# EEG-based Classification of Drivers Attention using Convolutional Neural Network

Fred Atilla
*Department of Cognitive Science and Artificial Intelligence*
*Tilburg University*
Tilburg, Netherlands
f.atilla@tilburguniversity.edu

Maryam Alimardani
*Department of Cognitive Science and Artificial Intelligence*
*Tilburg University*
Tilburg, Netherlands
m.alimardani@tilburguniversity.edu

*Abstract*—Accurate detection of a driver's attention state can help develop assistive technologies that respond to unexpected hazards in real time and therefore improve road safety. This study compares the performance of several attention classifiers trained on participants' brain activity. Participants performed a driving task in an immersive simulator where the car randomly deviated from the cruising lane. They had to correct the deviation and their response time was considered as an indicator of attention level. Participants repeated the task in two sessions; in one session they received kinesthetic feedback and in another session no feedback. Using their EEG signals, we trained three attention classifiers; a support vector machine (SVM) using EEG spectral band powers, and a Convolutional Neural Network (CNN) using either spectral features or the raw EEG data. Our results indicated that the CNN model trained on raw EEG data obtained under kinesthetic feedback achieved the highest accuracy (89%). While using a participant's own brain activity to train the model resulted in the best performances, inter-subject transfer learning still performed high (75%), showing promise for calibration-free Brain-Computer Interface (BCI) systems. Our findings show that CNN and raw EEG signals can be employed for effective training of a passive BCI for real-time attention classification.

## I. Introduction

Driver's inattention and fatigue has been shown as one of the major causes of road accidents [1]. A possible solution for this issue is a brain-computer interface (BCI) system that could detect and prevent such inattentive behavior from electroencephalography (EEG) signals [1]. The reliability of this system will depend on its ability to be applicable in a real-world driving environment where there is a variety of sensory information. Studies attempting to develop attention detection systems have been conducted in various environments such as classrooms [2, 3], driving [4, 5], and flying [6]. While the terms used in these studies differ (e.g. inattention [3], drowsiness [5, 7], invigilance [6], mental fatigue [8], mind-wandering [9], etc.), they all refer to states of reduced attention in an interchangeable way.

Initial work mainly applied Machine Learning (ML) models on users' brain activity to classify their attention state [10]. Alirezaei and Sardouie [2] classified attention states using EEG signals recorded in educational environments. After extracting the most efficient features (related to the beta band), they achieved 92% accuracy with Support Vector Machines (SVM) and Linear Discriminant Analysis (LDA) classifiers. Rundo et al. [7] performed a drowsiness/wakeful states classification based on EEG data recorded in a controlled setting and achieved 98% accuracy using an SVM algorithm. In order to improve ecological validity, some studies evaluated ML classification performances when EEG data were recorded in a driving simulator [5]. In [5], classification was based on power spectral density feature extraction and achieved up to 83% accuracy with the most efficient algorithm.

Amongst the many measures that can be used to characterize inattentive states in drivers, EEG is often used and achieves accuracies similar to other behavioural and physiological measures [10]. Significant neural correlates have been associated with inattention in sustained-attention driving tasks. Namely, drowsiness was found to be reflected by tonic alpha-band power increases in occipital and parietal regions along with phasic alpha power fluctuations [11]. It has also been found that as driving fatigue increases, theta band powers increase [12]. Finally, beta waves (characteristic of alertness and arousal) reduction is also known to indicate drowsiness states [3]. As attention is reflected in these frequency bands, current systems often perform spectral feature extraction from raw EEG signals before training a model. These methods suffer from time-consuming pre-processing as well as loss of information and ultimately depend on the researcher's expertise to select the relevant features [13].

Recent studies have tried to make up for these shortcomings by employing Deep Learning (DL) models such as convolutional neural networks (CNN). These models can extract the ideal features from raw EEG signals on their own and facilitate end-to-end learning [14, 15]. CNN have been used for drowsiness detection in different settings. Hajinoroozi et al. [4] compared channel-wise CNN (CCNN) with other deep learning (DNN, DBN) and machine learning methods (SVM, LDA), and obtained the highest accuracy (83%) with CCNN on a driver state (alert vs. drowsy) prediction task. Similar raw EEG-based predications were implemented by [9] for mind-wandering vs. alertness classification on a repetitive "backward counting" task and reached 92% accuracy. However, the high accuracies in [9] were obtained from EEG signals of only two participants, recorded in a screen-based task with little environmental noise and sensory feedback. These additional factors are inherent to real-world driving environments, therefore studies developing real-time attention detection systems should account for variability between sessions and environments just as they account for variability between individuals.

In real-world environments, drivers receive a variety of sensory information, which impacts their attention and brain activity. Chuang et al. [16] demonstrated that EEG signals during "poor performance" driving showed less significant increase in theta activity in the central and frontal areas when they were recorded in a static (visual input only) rather than in a dynamic driving simulator (visual and kinesthetic feedback). According to [17], providing participants with additional kinesthetic feedback results in a shift in the brain activity from the Task Positive Network (TPN) to the Default Mode Network (DMN). TPS is responsible for processing of sensory-perceptual information from the external world and DMN is known to reflect internally driven cognitive processes. The shift from TPN to DMN is behaviorally reflected by higher mind-wandering rates as well as less drowsiness due to lower perceptual demand. Therefore, additional sensory feedback in real-world environments can cause changes in EEG activity, reflecting different brain network dynamics [17].

In addition to being operational in a dynamic environment, an attention state classifier must be reliable for every user despite the high subjective variability in EEG patterns associated with perceptual decoupling and drowsiness [18]. Previous BCI research has mostly relied on subject-specific classifiers in which a model trained on one specific individual is used to make prediction about the same person. As calibrating the BCI for each user requires large amount of data and time, a subject-independent classifier in which the data of many users is used for model training could be more practical [19]. Examples of subject-independent classification are "mixed-subject approach" in which data from all subjects are mixed and split for training and test, and "inter-subject approach" in which a subject is left out for testing and the rest of subjects are used for training of the model [8, 18].

Considering all these challenges, this paper compares the performance of a state-of-the-art CNN classifier to SVM in a realistic simulated driving setting. In addition to the role of EEG activity type (raw signal vs. extracted spectral features), we investigated the impact of kinesthetic feedback (with vs. without feedback) and model training approach (mixed-subject vs. inter-subject) on model performances. We hypothesized that the CNN model would achieve better performances compared to the SVM model when it extracts task-relevant features from raw EEG signals on its own. Additionally, we expected that providing sensory feedback would reduce the classifier's performance by introducing more noise and inattention components in the EEG signals. Finally, given the variation in individuals' brain activity, we assumed that a mixed-subject approach would show higher classification accuracies than an inter-subject approach in all three models, but that the CNN model would be able to outperform SVM in the inter-subject approach.

## II. METHOD

### A. Dataset Description and Preparation

The dataset used in this experiment originated from [20]. The dataset included 14 subjects who performed a driving task in an immersive simulator. The car randomly deviated from the cruising lane to which subjects must react and correct the deviation (Fig. 1a). The reaction time (RT) in each deviation was used to determine driver's state as either *attentive* or *inattentive* in that trial. The task was repeated in two sessions; in one session subjects received realistic kinesthetic feedback from the car simulator (K+) and in the other they did not (K-). In both sessions, EEG signals were recorded from 30 electrodes at a sampling rate of 500 Hz (Fig. 1b).

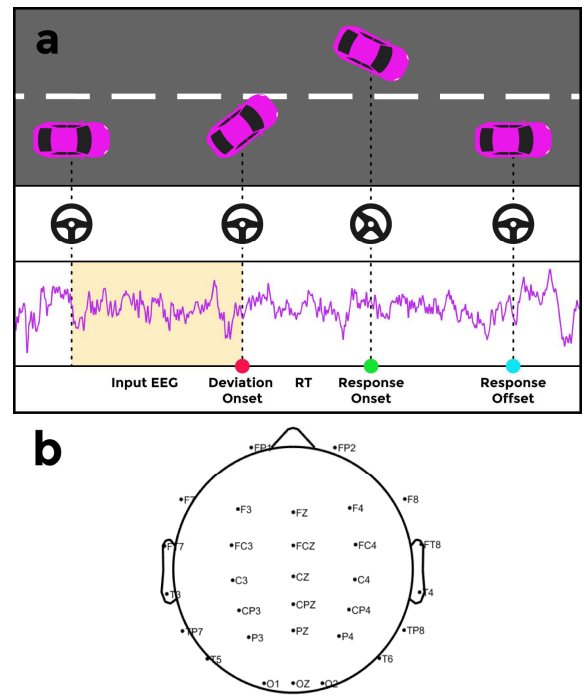

Fig. 1. Experimental design. (a) Subjects drove a car in a virtual simulator while their brain activity was recorded. The car randomly deviated from the lane (deviation onset) and subjects had to correct the deviation as fast as possible by steering back onto the correct lane (response onset). Their reaction time (RT) was used to label the subject state as attentive or inattentive. (b) Layout of the 30 electrode channels.

The dataset was already preprocessed with a 1-50 Hz band-pass filter, and by removing artefacts. Each session was marked with the timing of deviations (deviation onset) and responses (response onset/offset). Similar to the original study [20], the 20th percentile of RT values in each session was selected as the threshold for alertness. A session's 20th percentile RT represented an alert RT and any deviation trial with an RT faster than this was labeled as attentive. Based on the RT distributions, any trial with an RT slower than 2.1 sec was labeled as inattentive. This categorization excluded moderate reaction times that could be attributed to other cognitive states than low or high attention. EEG activity in a 3-sec window before the deviation was chosen as classifier input for attention state detection (Fig. 1a).

### B. Models and Input Data

A deep learning and a machine learning model were compared in this study. First, an established CNN called EEGNet [21] was trained in two ways; i) raw EEG data in the 3-sec window was directly fed to the model as input and ii) using discrete Fast Fourier Transform, mean spectral band powers were extracted from the EEG signals and used as model input. The frequency bands were defined as the delta (0-4 Hz), theta (4-7 Hz), alpha (8-12 Hz), beta (12-30 Hz), and gamma (30+ Hz) bands. Second, an SVM was trained on only these five frequency band powers.

The EEGNet was tuned with random search cross-validation over 5 folds and 60 epochs with a batch size of 32 to find the best hyperparameters. An Adam optimizer with a learning rate of 0.0001 was used to minimize the binary cross-entropy loss. The main layers and their tuned parameters are shown in Table 1. Any parameter not mentioned was kept as the default value in the original EEGNet architecture [21].

TABLE I. PARAMETERS IN EEGNET ARCHITECTURE. K = NUMBER OF KERNELS, F = SIZE OF KERNELS, D = DEPTH MULTIPLIER, N = NUMBER OF UNITS, A = ACTIVATION FUNCTION, P = PROBABILITY RATE.

| Block | Layers | Parameters | Output |
|---|---|---|---|
| 1 | Input |  | (30, 1501, 1) |
|  | Conv2D | K = 32, F =1x128 | (30, 1501, 32) |
|  | BatchNorm2D |  | (30, 1501, 32) |
|  | Depth-Conv2D | F = 1501x1, D = 4, A = ELU | (1, 1501, 128) |
|  | BatchNorm2D |  | (1, 1501, 128) |
|  | AvgPool2D | F = 1x16 | (1, 94, 128) |
|  | Dropout | P = 0.5 | (1, 94, 128) |
| 2 | Sep-Conv2D | K = 32, F = 1x16, A = ELU | (1, 94, 32) |
|  | BatchNorm2D |  | (1, 94, 32) |
|  | AvgPool2D | F = 1x8 | (1, 12, 32) |
|  | Dropout | P = 0.5 | (1, 12, 32) |
| 3 | Flatten |  | (384) |
|  | Dense | N = 1, A = Sigmoid | (1) |

## C. Evaluation

Both CNN and SVM models were trained and tested using two approaches. In one approach, samples from all subjects were mixed together and used to train the model (mixed-subject approach). As the models need the same number of samples for class, we identified the class with the least samples and randomly removed samples from the other class. The final dataset contained 769 attentive and 769 inattentive EEG segments, resulting in a tensor of 1538 (segments) × 30 (channels) × 1501 (data points). This dataset was split into 90% for training and 10% for testing.

In the other approach, samples from each subject were kept separate and used in leave-one-subject-out cross validation (inter-subject approach). Over 14 folds, the models were trained on the data of 13 subjects and tested on the left out subject. The accuracies from 14 iterations were then averaged for a final accuracy. This method is akin to inter-subject transfer learning and has been shown to perform well before [8].

## III. RESULTS

Table 2 gives all model accuracies for the different training approaches and feedback conditions. The highest accuracy (88.96%) was achieved by EEGNet trained on mixed-subject raw EEG with kinesthetic feedback. Overall, training the EEGNet on raw data resulted in better accuracies than training it on frequency bands. Comparison of the two models (EEGNet vs. SVM) revealed an effect of feedback; in both mixed- and inter-subject approaches, EEGNet trained with raw EEG performed better than SVM when kinesthetic feedback was present (K+ condition), whereas in the absence of feedback (K- condition) SVM performed better.

To further investigate the impact of kinesthetic feedback on the performance of attention classifiers, the accuracies obtained from all 14 subjects in the inter-subject approach were statistically compared. Fig. 2 presents boxplots of SVM and EEGNet performances in each feedback condition. A Wilcoxon Signed-Ranks test indicated that in the K- condition, SVM performed significantly better than EEGNet ($Z = -2.040$, $p = 0.041$), whereas in the K+ condition, the same test could not confirm a significant difference between the model peformances ($Z = -1.664$, $p = 0.096$).

TABLE II. ACCURACIES OF THE DEEP LEARNING (EEGNET) AND MACHINE LEARNING (SVM) MODELS TRAINED ON MIXED-SUBJECT AND INTER-SUBJECT DATA WHILE SUBJECTS RECEIVED EITHER KINESTHETIC FEEDBACK (K+) OR NO FEEDBACK (K-).

| Model | Mixed-Subject | | Inter-Subject | |
|---|---|---|---|---|
|  | *K+* | *K-* | *K+* | *K-* |
| SVM on EEG spectral features | 82.46 | 85.71 | 69.73 | 77.20 |
| EEGNet on EEG spectral features | 83.12 | 80.52 | 68.80 | 71.75 |
| EEGNet on raw EEG | **88.96** | 81.82 | 75.51 | 69.35 |

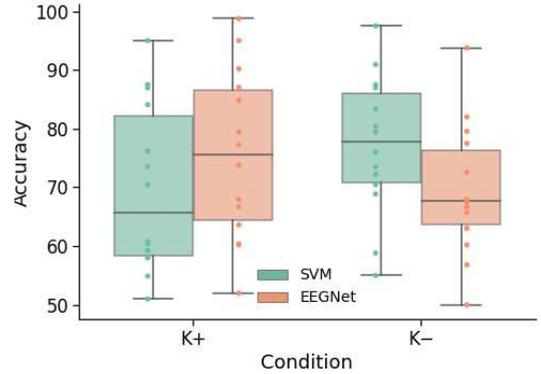

Fig. 2. Classification accuracies obtained by the EEGNet and SVM models in the inter-subject approach with kinesthetic feedback (K+) and without it (K-).

As can be seen in Fig. 2, the classification performance varied significantly across the participants with accuracies ranging from 50% to almost 100%. To investigate a possible reason for the high variation among subjects' performances, a Pearson correlation analysis between subject's reaction time dispersion (standard deviation of RT) and model accuracy was conducted in each condition, however it did not show a significant correlation for the K+ condition ($p = 0.629$) and neither for the K- condition ($p = 0.890$).

## IV. DISCUSSION

This paper examined the use of a convolutional neural network trained on raw EEG signals to classify driver's attention. A CNN model called EEGNet [21] was trained on either raw EEG signals or extracted spectral features and was compared against an SVM model only trained with EEG features. The results confirmed our hypothesis in that EEGNet reached the highest accuracy (89%) when it was trained on raw EEG in the mixed-subject approach and in the presence of kinesthetic feedback. Moreover, both SVM and EEGNet models reached an acceptable accuracy in the inter-subject approach, with EEGNet showing slightly higher performance when additional sensory feedback from the immersive simulator was present.

Past studies had already shown that CNN models are effective feature extractors, which can facilitate end-to-end learning with raw EEG signals [13, 15]. This is particularly handy in operational environments (e.g. driving) where the impact of environmental noise and perceptual demand is reflected on EEG activity [16, 17]. To examine the validity of our attention classifiers in real-world operation, we trained the models in two conditions; with and without kinesthetic feedback. Our results showed that in the presence of kinesthetic feedback, CNN could extract relevant features better from the complex EEG data, whereas SVM performed better with EEG features underlying simpler events.

Additionally, the mixed-subject approach always led to better classification performance than inter-subject approach. This was not surprising, as the classifier had possibly seen data from all subjects during the training phase and learned subject-specific patterns. Nevertheless, the inter-subject approach achieved a maximum accuracy of 77% using SVM. This is similar to state-of-the-art reports that applied inter-subject transfer learning [8, 22], suggesting the possibility of a universal attention BCI that does not require calibration for new users.

Kalaganis et al. [23] found a significant correlation between the achieved classification accuracy by the subjects and the difference in the response times they showed in attentive vs. inattentive states. In other words, the larger the gap between the reaction times in the two states, the higher accuracy the classifier could achieve. Unlike [23], we could not confirm this theory as correlations between subject accuracies and the spread of their reaction times were non-significant.

Several limitations should be mentioned. The dataset was small making the generalizability of the results outside the training data questionable. Future studies can employ larger datasets with more samples that would improve robustness of the classifier predictions. Another limitation is that the models were only trained on the fastest and slowest reaction times as the threshholds were defined by the researchers. This removed many EEG segments from the analysis as they were associated with neither attentive nor inattentive states. Future research should attempt to examine the performance of deep learning models in real-time driving or in Virtual Reality simulations where richer feedback is present [24]. This will further shed light on the effect of sensory feedback on brain activity and attention classifiers.

In conclusion, convolutional neural networks are promising feature extractors for classification of attention state from raw EEG signals during demanding driving tasks. Moreover, our results show that data from multiple subjects can be used to train a subject-independent classifier, saving time and resources on data collection and model calibration. In practice, the present findings can be useful in the development of future driver assistance systems, leading to the reduction of traffic accidents and improvement of road safety.